\documentclass[reprint, aps, pra, twoside, floatfix, nofootinbib, longbibliography]{revtex4-2}
\usepackage{amsmath, amssymb}
\usepackage{xfrac}
\usepackage{hyperref}
\hypersetup{colorlinks=true, linkcolor = blue, citecolor = blue, urlcolor = blue}
\usepackage{graphicx}
\usepackage{epstopdf}
\usepackage{times}
\usepackage[caption=false]{subfig}
\usepackage{tikz}
\begin{document}
	\title{Laser beam fluctuations for long distance propagation in the turbulent atmosphere}
	\author{V~Andriichuk}
	\email{valentyn.andriichuk@gmail.com}
	\author{O~Chumak}
	\email{chumak@iop.kiev.ua}
	\author{L~Derzhypolska}
	\author{I~Matsniev}
	\affiliation{Institute of Physics, National Academy of Sciences of Ukraine, \\ pr.~Nauky 46, 03028 Kyiv, Ukraine}
	
\begin{abstract}
	The method of photon distribution function (PDF) is used to study fluctuations of light beams propagating through a turbulent atmosphere. Our analysis concerns the regime of saturated fluctuations. The focus is on the phenomena of beam wandering and the effect of partial coherence on photon density fluctuations. It is shown that the size of the quasiclassical part of wandering decreases with the propagation distance, while the quantum part increases. We explain this qualitative difference by beam fragmentation, which is accompanied by a loss of correlation between individual parts.
		 
	The effect of the phase diffuser on the fourth moment (FM) of the irradiance is taken into account. The obtained explicit expression for the FM indicates the possibility of a significant reduction of noise in the communication channel. The diffuser changes the shot noise from delta-correlated (in the spatial domain) to smoothly distributed. The theory developed here is used to estimate the influence of the phase diffuser on light fluctuations.
\end{abstract}
\maketitle
\section{Introduction}
	For many years, theoretical and experimental studies of light propagation in the Earth's atmosphere were stimulated by practical needs \cite{tat, fan1, fan2, ric, and, sem1, sem2, usen}. Understanding the physical mechanisms of noise generation in laser beams caused by random inhomogeneities of the atmosphere can help build reliable communication systems. 
	Many characteristics of the irradiance are described by the photon distribution function (PDF), which is the density of photons in the coordinate-momentum (phase) space. This function is a quadratic form of light amplitudes with various wave vectors \cite{rar,sus,ber}. The PDF method is effective not only in atmospheric optics but also for studying the propagation of electromagnetic quanta in one-dimensional waveguides \cite{st1, st2}.

	Linear kinetic equations describe both the evolution of the mean and fluctuating components of the PDF. They can be easily solved using Fourier transforms. Then the densities (or fluxes) of photons in the configuration domain are obtained by integration over wave vectors. 

	Noise characteristics of light can be described by the mean square (MS) of fluctuations of photon density. Direct obtaining of these characteristics leads to a multiple integral that can not be evaluated analytically. Numerical studies also seem problematic. The situation is not significantly simplified even in the case of paraxial beams.

	The problem can be solved for cases of weak and strong turbulence. In the paper \cite{b}, an iterative procedure has been proposed and successfully applied to systems with weak or even moderate turbulence. For the case of strong turbulence \cite {bb}, a completely different approach was implemented. This approach uses modifications of the statistical properties of the beam over the long path of propagation. The modifications are due to multiple photon scattering by atmospheric inhomogeneities. These inhomogeneities are caused mainly by atmospheric vortices. 

	This paper is a continuation of Ref.~\cite{bb}. Using the results of \cite{bb}, we obtain the long-distance classical and  quantum components of the  beam wandering. We show that the corresponding characteristic lengths have very different time dependences. This important property of the irradiance should be considered not only for numerical research but also for simple estimates.
 
	The influence of partial coherence on the fourth-order correlation function is also investigated. The general expression, obtained here, qualitatively differs from that in \cite {bb}. The main difference is in the quantum part of the correlation function. Due to the finite responce time of receiver device and the presence of the phase diffuser, the shot-noise term is no longer delta-correlated. The spatial distribution of this noise is described by a finite-width distribution, i.e. different from that in \cite{bb}. 

	The modification of the quasiclassical component due to the phase diffuser can be very significant. In the specific case of the coincidence of coordinates in the FM, it can be used to obtain the scintillation index, which describes the quality of the atmospheric channel. 

	The rest of this paper is organized as follows. 
	Section \ref{sec:Kin} presents the already known results, which are used in the following parts to obtain new ones. Section~\ref{sec:Wander} illustrates the analytical derivation of the wandering length. The derivation of the general expression describing the influence of the phase diffuser on the fourth-order correlation function is given in Sec.~\ref{sec:Dif}. The discussion of the obtained results is in Sec.~\ref{sec:Dis}.
\setlength{\unitlength}{1cm}

\begin{figure}[t!]
	\centering
	\begin{tikzpicture}
		\node[above right,inner sep=0] (image) at (0,0) {
			\includegraphics[width=0.49\textwidth]{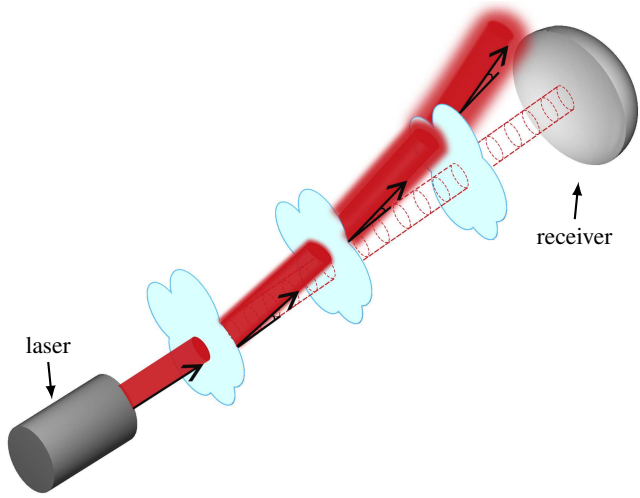}
		};
		\draw[latex-, thick, black] (0.75, 1.35) -- ++(-0.05, 0.50) node[above, black]{laser};
		\draw[latex-, thick, black] (7.75, 4.25) -- ++(-0.05, -0.50) node[below,black]{receiver};		
	\end{tikzpicture}
	\caption{A schematic representation of laser beam propagation through a turbulent atmosphere over a long distance. The beam diameter after leaving the laser aperture is less than the characteristic size of atmosphere inhomogeneities. This consideration is legitimate because, at the initial moment, the probability of collision with larger size inhomogeneities is greater than vice versa. Thus, there is the beam deflection as a whole from the aperture of the receiver. Nevertheless, due to the strong turbulence of the atmosphere, there is also a little beam boarding.
	\label{fig:fig_1}}	
\end{figure}
\section{Kinetic equation for photon density  in the phase space}
\label{sec:Kin}
	The photon distribution function is defined by analogy with distribution functions, which are widely used in solid state physics. These can be distributions for electrons, phonons, etc. (See, for example, \cite {kog,tar}). In our case, the photon distribution function ({PDF}) is defined as
\begin{equation}
f({\bf r},{\bf q},t)=\frac 1V\sum_{\bf k}e^{-i{\bf k\cdot r}}b^\dag_{{\bf
		q}+ {\bf k}/2}b_{{\bf q}-{\bf k}/2},
	\label{odyn}
\end{equation}
	where $b^\dag_{\bf q}$ and $b_{\bf q}$ are creation and annihilation operators of photons with the wave vector ${\bf q}$; $V\equiv L_xL_yL_z\equiv SL_z$ is the normalizing volume. This function describes the spatial distribution of light in a beam with an accuracy of several wavelengths. The limitation of the accuracy of this approach is caused by the coarsening procedure for the PDF, which is typical for the kinetics of multiparticle systems (see, for example, \cite{sus,ber,man, kol}). The distribution function can be represented by the sum of the mean and fluctuation parts
	\begin{equation}
		f=\langle f \rangle + \delta f,	
		\label{dva}
	\end{equation}
	where the symbol $\langle \cdot \rangle$ means both quantum mechanical and statistical averaging. The kinetic equations describing photon evolution are given by
\begin{equation}
	\bigg(\partial_t +{\bf c_q}\cdot\partial_{\bf r}+\hat{\nu}_{\bf q}\bigg) \langle f({\bf r},{\bf q},t)\rangle=0,
	\label{try}
\end{equation}
\begin{equation}
	\bigg(\partial_t +{\bf c_q}\cdot\partial_{\bf r}+\hat{\nu}_{\bf q}\bigg) \delta f({\bf r},{\bf q},t)=K_L({\bf r},{\bf q},t),
	\label{chotyry}
\end{equation} 
	where ${\bf c_q}=\frac{\partial\omega_{\bf q}}{\partial{\bf q}}$, 
	$\hbar\omega_{\bf q}=\hbar cq$ are the photon velocity and photon energy, respectively; $\hat{\nu}_{\bf q}$ is the operator describing \textquotedblleft collisions\textquotedblright\,of photons with atmospheric inhomogeneities (collision integral). The characteristic size of inhomogeneities is considered to be much larger than the wave length of the irradiance. The term $ K_L({\bf r},{\bf q},t)$ is the Langevin \textquotedblleft source\textquotedblright\,of fluctuations, which average value, $\langle K_L\rangle $, is equal to zero. The \textquotedblleft source\textquotedblright\,term can be interpreted as fluctuations in the number of collision events per unit time. The correlation function of the \textquotedblleft source\textquotedblright\,and the collision term are expressed through the probabilities of collision events and provide a realistic description of the influence of the atmosphere on photon kinetics.
	The explicit form of the collision integral for paraxial beams propagating in the z-direction is given by
\begin{gather}
	\hat{\nu}_{\bf q}\big \{ f({\bf r},{\bf q},t)\}=\frac{2\pi\omega_{0}^{2}}{c}\int d{\bf k'_{\bot}}\psi({\bf k'_{\bot}})\big(f({\bf r},{\bf q},t) \nonumber \\ -f({\bf r},{\bf q+k'_{\bot}},t)\big).
	\label{pyat}
\end{gather}

	The notation $(_\bot)$ denotes the components of the corresponding vector, orthogonal to the z-axis, and  $\psi ({\bf k'_{\bot}})=\frac V{(2\pi)^3}|n_{\bf k'_{\bot}}|^2$, where $n_{\bf k}$ is the Fourier transform of the random part of the  refraction index  defined by
\begin{equation}\label{shist}
	n_{\bf k}=\frac 1V\int dVe^{i{\bf k\cdot r}} n({\bf r}).
\end{equation} 
	The collision integral represents a regular part of collision processes. It follows from \eqref{pyat} that only the transversal components of the wave vectors can be changed due to collisions. This occurs even in the case of a statistically homogeneous and isotropic atmosphere. The paradox can be explained by the relativistic contraction of the lengths of moving objects. The contraction is due to the relative motion of photons and inhomogeneities in the direction of beam propagation. It is manifested here as a strong artificial anisotropy of the atmosphere. 

	Equations \eqref{try} and \eqref{chotyry} are obtained in \cite{b} using the Heisenberg representation for the PDF and the  Hamiltonian of photons in a turbulent atmosphere. The latter is given by  
\begin{equation}\label{sim}
	H=\sum_{\bf q}\hbar\omega_{\bf q}b^\dag_{\bf q}b_{\bf q}-\sum_{\bf q,k}\hbar\omega_{\bf q}n_{\bf k}b^\dag_{\bf q}b_{\bf q+k}.
\end{equation}
 	In the case of paraxial beams, the coordinate $z$ in Eq.~\eqref{odyn} can be replaced by the distance traveled by photons, $ct$, after  leaving the aperture plane. Moreover, the paraxial regime of propagation means that the component of the wave vector $q_z $  in \eqref{try} and \eqref{chotyry} can be replaced by $q_0$. Thus, these equations describe the two-dimensional picture. We use here the solution for the average PDF in the plane orthogonal to the $z$-axis \cite{bb}. It is given by
\begin{equation}
	\begin{split}
		\langle f({\bf r},{\bf q},t)\rangle {=} & \, C\int d{\bf p}\int d{\bf k}e^{-i{\bf k} \cdot{\bf r}-i{\bf q}\cdot({\bf p}-{\bf k} ct/q_o)} \\
		& \, \times e^{- k^2r_0^2/8-p^2/2r_0^2}e^{-\int\limits_0^tdt'\gamma({\bf k},{\bf p},t')}.
	\end{split}
	\label{visim}
\end{equation}
	All vectors here are orthogonal to the $z$-axis. The notation ($ \bot $) is omitted for the sake of bravity. The constant $C$ can be expressed in terms of the total flux of photons. The equation \eqref{visim} was obtained  assuming the Gaussian profile, $e^{-r^2/{r_0^2}}$, of the laser field in the aperture plane. The parameter of this profile, $r_0$, is included in \eqref{visim} explicitly. The last factor under the integral in \eqref{visim}, where     
\begin{equation}
	\gamma({\bf k},{\bf p},t)=\frac{4\pi\omega_{0}^{2}}{c}\int d{\bf k'}\psi({\bf k'}) \sin^2\bigg[\bigg({\bf p}-{\bf k}\frac{ct}{q_0}\bigg){\cdot} 
	\frac{\bf k'}2\bigg],
	\label{devyat} 
\end{equation}
	describes the effect of collisions throughout the propagation time $t$. The integration in \eqref{visim} and \eqref{devyat} includes many variables, namely ${\bf k, k^{'}, p}, t^{'}$. 
	The analysis is facilitated in the case of long propagation time when   
\begin{equation}\label{desyat} 
	\nu t\gg1,
\end{equation} 
	were $\nu=\frac{2\pi\omega_0^2}c\int d{\bf k}\psi({\bf k})$. The condition \eqref{desyat} implies that there are many collisions of photons during the propagation time $t$. In this case, only small arguments of the function $\sin^2\big[\big({\bf p}-{\bf k}\frac{ct}{q_0}\big){\cdot} \frac{\bf k'}2\big]$ make a significant contribution to $\gamma({\bf k},{\bf p},t)$.   
	Thus, one can use the approximation $\sin^2 x \approx x^2$. 
	As a result, many integrations in \eqref{visim} can be performed analytically and we have 
	\begin{equation}
		\begin{split}
			\langle f({\bf r},{\bf q},t)\rangle = & \, 3C\bigg(\frac{2\pi q_0}{ct^2\alpha}\bigg)^2 \\ & \, \times\exp{\left[{-}\left({\bf r}{-}\frac{{\bf q}ct}{2q_0}\right)^2\frac 4{\langle {\bf r}^2\rangle_T}{-}\frac{4{q}^2}{\langle {\bf q}^2\rangle_T}\right]}.
			\label{odynad} 
		\end{split}
	\end{equation}
 	The constant $\alpha$ is given by
\begin{equation}
	\alpha =\frac{\pi\omega_0^2}{2c}\int d{\bf k}^\prime  k^{\prime 2}\psi(k^\prime).
	\label{dvanad}
\end{equation}

	The values
	$\langle  {\bf r}^2\rangle={4z^3c\alpha}/({3\omega_0^2})$ and $\langle {\bf q}^2\rangle=4\alpha t$ describe  the increase of beam radius and photon momentum caused by \textquotedblleft collisions\textquotedblright, respectively.

	Equation \eqref{odynad} does not  depend on the value $r_0$, which can be neglected in the case of long  propagation time [see inequality \eqref{desyat}]. It corresponds to the physical picture, where the beam broadening is much larger than $r_0$. 
 
	The expression \eqref{odynad} can be used to describe the parameters of light beams at any time $t$, which satisfies inequality \eqref{desyat}. It can also be used directly to obtain the FM of the light field. 
	The condition \eqref{desyat} means that the Gaussian statistics can be applied to describe  photon fluctuations. In this case, the FM of the light field can be expressed by \cite{bb}
\begin{multline}
	\langle\delta\hat{I}({\bf r}, t)\delta\hat{I}({\bf r^\prime}, t)\rangle=\delta({\bf r}-{\bf r^\prime})\langle\hat{I}({\bf r}, t)\rangle \\
	+\sum_{{\bf q},{\bf q^\prime}}\langle f(\frac{{\bf r}+{\bf r^\prime}}{2},{\bf q})\rangle\langle f(\frac{{\bf r}+{\bf r^\prime}}{2},{\bf q^\prime})\rangle e^{i({\bf q^\prime}-{\bf q})\cdot({\bf r}-{\bf r^\prime})},
	\label{trynad}
\end{multline}
	where $\langle\hat{I}({\bf r}, t)\rangle=\sum_{{\bf q}}\langle f({\bf r} ,{\bf q},t)\rangle$ and  $\delta\hat{I}({\bf r}, t)=\sum_{{\bf q}}\delta f({\bf r} ,{\bf q},t)$ are the mean and fluctuating parts of the spatial photon density, respectively. 
	The characteristic values of $q$ and $q^\prime$ are determined by the value $\langle q^2\rangle^{1/2}$. Then from the right part of \eqref{trynad}, it follows that the characteristic distance $|{\bf r}-{\bf r^\prime}|$, given by $\langle q^2\rangle^{-1/2}$, is the correlation length of the irradiance. 

	\begin{figure}[t!]
		\centering
		\begin{tikzpicture}
			\node[above right,inner sep=0] (image) at (0,0) {
				\includegraphics[width=0.47\textwidth]{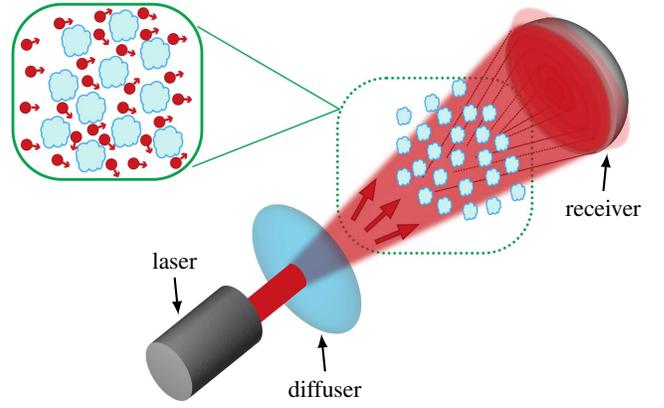}
			};
			\draw[latex-, thick, black] (2.30, 1.30) -- ++(-0.05, 0.50) node[above, black]{laser};
			\draw[latex-, thick, black] (4.20, 1.05) -- ++(0.05, -0.50) node[below, black]{diffuser};
			\draw[latex-, thick, black] (8.0, 3.45) -- ++(-0.05, -0.50) node[below,black]{receiver};
			
		\end{tikzpicture}
		\caption{A phase diffuser is added to the scheme presented in \figurename\ref{fig:fig_1}. Thus, the laser beam, after leaving the aperture of laser, passes through the phase diffuser with subsequent fragmentation (broadening) of the beam. The green rectangle shows that the laser beam \textquotedblleft fragments\textquotedblright \, have various propagation directions (so, the laser beam broads). And the characteristic sizes of inhomogeneities are larger than the diameters of beam \textquotedblleft fragments\textquotedblright. Consequently, diffraction of these \textquotedblleft fragments\textquotedblright on small atmospheric inhomogeneities is observed. As a result, over a long propagation distance, the laser signal is amplified in some directions, which is detected in the receiver.
			\label{fig:fig_2}}
	\end{figure}

	If we ignore the contribution of the shot noise given by first term on the right side of Eq.~\eqref{trynad} and consider the case ${\bf r}={\bf r^\prime}$, we obtain the scintillation index $\sigma^2=\langle(\delta I({\bf r},t))^2\rangle/\langle I({\bf r},t)\rangle^2$ equal to one. This result, known in the literature (see \cite{fan1,kra}), is immediately obtained from Eq.~\eqref{trynad}. 
 
	Usually the shot noise is created by classical particles, which in the course of movement can randomly overcome or be scattered by potential barriers. Since we deal with waves of light, the first term on the right side of \eqref{trynad} can be interpreted as a manifestation of the discreteness of the light field. In other words, this term suggests that the light field, in addition to the properties of classical waves, also has the properties of particles, i.e. quantum properties.

	The second term on the right side of \eqref{trynad} can be interpreted as the result of interference of four classical waves. Its explicit form is obtained after substituting \eqref{odynad} into \eqref{trynad} and integration over ${\bf q}$ and ${\bf q^\prime}$. The total correlation function of photon density is  expressed by 
\begin{gather}
	\langle\delta\hat{I}({\bf r},t)\delta\hat{I}({\bf r'}, t)\rangle=\delta({\bf r}-{\bf r'})\langle\hat{I}({\bf r}, t)\rangle+
	\left(\frac{4\pi SC}{\langle{\bf r}^2\rangle}\right)^2 \nonumber \\ \times \exp\left\{{-}\frac{({\bf r}{+}{\bf r'})^2}{2\langle{\bf r}^2\rangle}{-}\frac{({\bf r}{-}{\bf r'})^2\langle{\bf q}^2\rangle}{8}\right\},
	\label{chotyrn}
\end{gather}
	where  
\begin{equation}\label{pyatna}
	\langle\hat{I}({\bf r},t)\rangle=\frac{4\pi SC}{\langle{\bf r}^2\rangle}e^{{-r^2}/\langle{{\bf r}^2\rangle}}.
\end{equation} 
	The quantity $\langle\hat{I}({\bf r},t)\rangle$ is a two-dimensional photon density. Its integration  over ${\bf r}$ gives the total number of such photons, $N$. Then the ratio $4\pi SC/\langle{\bf r}^2\rangle$ is equal to
	$N/(\pi\langle r^2\rangle)=1/a^2$, where ${a^2}$ can be interpreted as the area per particle. 

	In the next section, we use equation \eqref{chotyrn} to obtain  the classical value of the wandering radius. A comparison will be made with a similar value derived from the shot-noise term. 

\section{Beam wandering: a comparison of classical and quantum terms}
	\label{sec:Wander}
	The position of beam centroid, ${\bf R}_w(t)$, is defined as
	\begin{equation}\label{shisnad}
		{\bf R}_w(t)=\frac {\int d{\bf r} {\bf r} I({\bf r},t)}{\int d{\bf r} \langle I({\bf r},t)\rangle }, 
	\end{equation}
	where $\int d{\bf r} \langle I({\bf r},t)\rangle=\pi\langle r^2\rangle/a^2.$
	In the case of symmetric beams, you can just write $\langle{\bf R}_w(t)\rangle=0$. The fluctuation part is described by its mean-square  
	\begin{equation}\label{simnad}
		\langle{\bf R}_w(t)^2\rangle=\frac {\int d{\bf r}d{\bf r}^\prime ({\bf r}\cdot{\bf r}^\prime)\langle\delta I({\bf r},t)\delta I({\bf r}^\prime,t)\rangle}{(\int d{\bf r} \langle I({\bf r},t)\rangle)^2}. 
	\end{equation}

	Like Eq.~\eqref{chotyrn}, the quantity \eqref{simnad} has also the quantum and classical parts: 
 		$\langle{\bf R}_w(t)^2\rangle=R^2_{sh}+R^2_{cl}$. The classical part is given by 
	\begin{equation}\label{visimn}
		R^2_{cl}=\frac 1{a^4}\frac{\int d{\bf r}d{\bf r}^\prime({\bf r}\cdot{\bf r}^\prime) \exp\left\{{-}\frac{({\bf r}{+}{\bf r^\prime})^2}{2\langle{\bf r}^2\rangle}{-}\frac{({\bf r}{-}{\bf r'})^2\langle{\bf q}^2\rangle}{8}\right\}}{(\int d{\bf r}\langle I({\bf r},t)\rangle)^2}. 
	\end{equation}
	Using the relation  $({\bf r}\cdot{\bf r}^\prime) = \frac 14[({\bf r}+{\bf r}^\prime)^2-({\bf r}-{\bf r}^\prime)^2] $  in Eq.~\eqref{visimn}, we get the difference of two integrals in which the negative term,   $-\frac14({\bf r}-{\bf r}^\prime)^2$, gives an insignificant contribution as compared to the contribution of the positive term, $\frac14({\bf r}+{\bf r}^\prime)^2$. The negative term can be neglected when
 	\begin{equation}\label{devyatn} 
		{\langle{\bf r}^2\rangle} \gg \frac{4}{\langle{\bf q}^2\rangle}.
	\end{equation}
	Inequality \eqref{devyatn} means that for long propagation times the radius of the beam, $\langle{\bf r}^2\rangle^{1/2}$, is much larger than the correlation distance $\langle{\bf q}^2\rangle^{-1/2}$. In this case, the expression for the classical value of the beam wandering is given by 
 	\begin{equation}\label{dvad}
 		R^2_{cl}(t)=\frac 2{\langle q^2\rangle}=\frac 1{2\alpha t}.
 	\end{equation}
	This expression shows that the classical mean-square, $ R^2_{cl}$, tends to zero with increasing propagation time as $1/t$. There is a simple explanation for the decay of wandering after long-distance propagation. The reason is the broadening of the beam that reduces the probability of being deflected as a whole (for a comparsion, see \figurename \ref{fig:fig_1} and \figurename \ref{fig:fig_2}). In short, the beam wandering at the beginning of the trajectory turns into the beam broadening at the end. This explanation is consistent with the opinion of R. Fante \cite{fan2} who considered that in strong turbulence diffraction dominates over refraction. 

	The quantum expression for $R^2_{sh}$ is significantly different from the classical expression given by Eq. \eqref{dvad}. Similar to the previous consideration, the explicit form of this term can be obtained from the general expression 
	\begin{equation}\label{dvad1}
		R^2_{sh}=\frac{\int d{\bf r}d{\bf r}^\prime({\bf r}\cdot{\bf r}^\prime) \delta({\bf r-r^\prime})\langle I({\bf r},t)\rangle}{(\int d{\bf r}\langle I({\bf r},t)\rangle)^2}. 
	\end{equation} 
	After simple integration, we get
	\begin{equation}\label{dvad2}
		R^2_{sh}(t)=\frac {a^2}\pi=\frac{\langle r^2\rangle}N=\frac{\langle r^2\rangle}{\int d{\bf r}{\langle I({\bf r},t)\rangle}}\sim t^3.
	\end{equation} 
	Here it is assumed that the total flux of photons in the $z$-direction is conserved. This condition is fulfilled in the case of paraxial beams. 

	Comparing the time dependences in the equations \eqref{dvad} and \eqref{dvad2}, we see a tendency of the  wandering to decrease with time in the region of dominant four-wave classical interference, $R^2_{cl}(t)$, while a pronounced increase of wandering, $R^2_{sh}(t)$, occurs in the region where quantum properties of radiation predominate. In the first case, the decay of wandering is due to an increase in the radius of the beam. In short, wandering at the beginning of the beam trajectory is replaced by beam broadening at the end. 

	In the second case, the fragmentation of the beams occurs, where the individual parts do not correlate with each other and make an independent contribution to the broadening. 

	In the next section, the approach based on Eq.~\eqref{trynad} is modified to account for the effect of the phase diffuser. 
 
\section{Influence of phase diffuser on long-distance fluctuations of photon density } 
	\label{sec:Dif}

	The theoretical approach based on the equation \eqref{trynad} should be modified for atmospheric  channels with a phase diffuser. Typically, the diffuser is used for partial randomization of the propagating field (see, for example, papers \cite{ric2, ba, qian, wang, gor}). It is placed  in front of the transmitter aperture and randomly varies the wavefront of the laser radiation. The diffuser causes the beam to spread beyond its diffraction size. At first glance, it seems that the phase diffuser only increases fluctuations in the channel. However, it increases characteristic wave vectors in the direction normal to the $z$-axis, and a reasonable choice of the diffuser can significantly reduce the scintillation index (see, for example, \cite{ric2}). 
	Our task here is to obtain the asymptotic expression for the correlation function of photon density similar to Eq.~\eqref {trynad}, which takes into account the combined effect of atmospheric turbulence and phase diffuser. 

	As before, the optical field in the aperture plane is described by the expression 
	\begin{equation}\label{dvad3}
		C_Lb(t)e^{-r^2/r_0^2},
	\end{equation} 
	where $b(t)=b_0e^{-i{\omega_0}t}$  is the amplitude of the generated irradiance, the constant $C_L$ is expressed by the average flux of irradiance in free space. The effect of the phase  diffuser will be taken into account by multiplying \eqref{dvad3} by $e^{-i{\phi}(\bf r)}$, where $\phi(\bf r)$ is the random phase that distorts the wavefront of the beam. The electromagnetic field generated by a laser  in free space is proportional to  
	\begin{equation}\label{dvad4}
		\sum_{\bf k}e^{i{\bf k\cdot r}}b_{\bf k}(t),
	\end{equation} 
	where all vectors are orthogonal to $z$-axis. Considering profiles of the two fields \eqref{dvad3} and \eqref{dvad4} to be identical  when $z=0, t=0$, we find 
	\begin{equation}\label{dvad5}
		b_{\bf k}(t=0)\sim\int d{\bf r} e^{-i({\bf k\cdot r}+\phi({\bf r}))-r^2/r_0^2}.
	\end{equation} 
	Similar to \eqref{dvad5}, we obtain the product of the amplitudes 
	\begin{gather}
		b^\dag_{{q}+\frac{\bf k}{2}}b_{{q}-\frac{\bf k}{2}}|_{t=0}\sim\int d{\bf r}d{\bf r}^\prime e^{i[({q}+\frac{\bf k}{2})\cdot{\bf r}-({q}-\frac{\bf k}{2})\cdot{\bf r}^\prime+\phi({\bf r})-\phi({\bf r}^\prime)]}\nonumber \\ \times e^{-(r^2+r^{\prime^2})/r_0^2},
		\label{dvad6} 
	\end{gather}
	which is part of PDF [see Eq.~\eqref{odyn}]. The random phase difference $\phi({\bf r})-\phi({\bf r}^\prime)$ in \eqref{dvad6} reduces the irradiance correlation if ${\bf r}\ne {\bf r}^\prime$. In the case of large characteristic values of $\phi$, only a small range of ${\bf r}-{\bf r}^\prime$ contributes significantly to the  integral in Eq.~\eqref{dvad6}. Therefore, it is reasonable to approximate $\phi({\bf r})-\phi({\bf r}^\prime)$ by a simpler function    
	\begin{equation}\label{dvad7}
		\frac{\partial\phi}{\partial{\bf r}}\cdot({\bf r}-{\bf r}^\prime)\equiv{\bf g}\cdot({\bf r}-{\bf r}^\prime),
	\end{equation}
	where it is assumed that a random quantity ${\bf g}$  satisfies the Gaussian (normal) statistics. In our case, the use of the Gaussian statistics can be justified by the presence of a finite response time of the detector. During this time, the field intensity is averaged. The Gaussian statistics is applicable if this time is much longer than the characteristic time of diffuser phase change (see, for example, works \cite{fan1,fan2}). This situation is known in the literature as a slow-detector case. In general, not only spatial but also temporal phase variations introduced by the phase diffuser should be considered. This case was analyzed in \cite{dyn}. It is convenient to use in Eq. \eqref{dvad6} new spatial variables
	\begin{equation}\label{dvad8}
		{\bf R}=\frac{\bf r+r^\prime}2, {\bf\Delta} ={\bf r}-{\bf r^\prime}. 
	\end{equation}
	This simplifies Eq.~\eqref{dvad6} to 
	\begin{equation}
		\begin{split}
		 b^\dag_{{q}+\frac{\bf k}{2}}b_{{q}-\frac{\bf k}{2}}|_{t=0} \sim & \, \int d{\bf R}e^{i{\bf k\cdot R}-2R^2/r_0^{2}} \\ & \, \times\int d{\bf \Delta} e^{i({\bf q+g)}\cdot{\bf \Delta}-\Delta^2/2r_0^2},
		\end{split}
		\label{dvad9} 
	\end{equation} 
	from which we see that the effect of phase diffuser is reduced to substitution of $\bf{q+ g}$ for $\bf {q}$ at $t=0$.
	After averaging  Eq.~\eqref{dvad9} over the Gaussian random quantity $ \bf {g}$, for which the relations  
	\begin{equation}\label{tryd}
		\langle e^{i[\phi({\bf r})-\phi({\bf r}^\prime)]}\rangle\approx \langle e^{i{{\bf g}\cdot{\bf \Delta}}}\rangle= e^{-\langle g^2\rangle\Delta^2/2}
	\end{equation}
	is applied, we can integrate over ${\bf R}$  and  ${\bf \Delta}$. [The relationships $\langle g_x^2\rangle=\langle g^2_y\rangle\equiv\langle g^2\rangle$ are assumed]. The result is
	\begin{equation}\label{tryd1}
		b^\dag_{{q}+\frac{\bf k}{2}}b_{{q}-\frac{\bf k}{2}}|_{t=0}\sim e^{-(q^2r_1^2+k^2r_0^2/4)/2},
	\end{equation}
	where $r_1^2=r_0^2/(1+ r_0^2\langle g^2\rangle)$. 

	It is seen from Eq.~\eqref{tryd1} that the characteristic value of $q$ at $t=0$ is of the order of  $ r_1^{-1}$  $( r_1^{-1}>r_0^{-1})$. This means that the receiver averages over a wider range of photon trajectories due to the presence of the phase diffuser. A similar effect can be achieved by increasing the receiver aperture. Therefore, the influence of the phase diffuser on the beam can be interpreted as an additional \textquotedblleft artificial\textquotedblright\,aperture averaging \cite{ric}. Enlargement of the receiver aperture is problematic in many practically important cases. A good alternative is to use a phase diffuser.  

	Using the expressions \eqref{odyn} and \eqref{dvad9}, we obtain the initial condition for the distribution function, which together with solution of the equation \eqref{try} provides the average PDF. This solution can be obtained using Eq.~\eqref{visim} by replacing there $p^2/(2r_0^2)$ with $p^2/(2r_1^2)$. 

	Without a diffuser, the asymptotic value of the fourth moment, which describes the fluctuations of the photon density, is expressed in terms of average PDFs [see Eq.~\eqref{trynad}]. Next, we will consider in detail a modification of this expression to take into account the effect of the phase diffuser. 

	The correlation of photon density at points ${\bf r}_A$ and ${\bf r}_B$ is described by
	\begin{gather}
		\langle\overline{\delta I({\bf r}_A,t) \delta I({\bf r}_B,t)}\rangle=\langle\overline{ I({\bf r}_A,t)I({\bf r}_B,t)}\rangle \nonumber \\ -\langle\overline{I({\bf r}_A,t)} \rangle\langle\overline{ I({\bf r}_B,t)}\rangle,
		\label{tryd2}
	\end{gather}
	\begin{equation}\label{tryd3}
		I({\bf r}_{A,B}, t)=\frac {1}{S}\sum_{{\bf q,k}} e^{-i{\bf k\cdot r}_{A,B}}b^\dag_{{\bf
		q}+ {\bf k}/2}b_{{\bf q}-{\bf k}/2}.
	\end{equation}
	The influence of the phase diffuser is taken into account in $\langle\overline{I({\bf r}_{A,B},t)}\rangle$ and $I({\bf r}_{A,B},t)$ by modifying the initial conditions as explained by equation \eqref{tryd1}. The first term on the right-hand side of equation \eqref{tryd2} is a linear form of the products of four operators at time $t$:
	\begin{equation}\label{tryd4}
		\langle \overline{b_ {{\bf q}{+}\frac{\bf k}{2}}^\dag b_ {{\bf q}{-}\frac{\bf k}{2}}b_ {{\bf q}_1{+}\frac{\bf k_1}{2}}^\dag b_ {{\bf q_1}{-}\frac{\bf k_1}{2}}}\rangle_t.
	\end{equation}
	The top line and angle brackets indicate two types of independent averages.  The first averaging occurs due to the the finite time of the responce of the measuring instrument (finite detection time). Equations \eqref{dvad6}-\eqref{dvad9} show that the combined effect of the \textquotedblleft slow detector\textquotedblright\,and the phase diffuser can be taken into account by replacing $e^{-q^2r_0^2/2-q^2_1r_0^2/2}$ with  $e^{-q^2r_1^2/2 -q^2_1r_1^2/2}$.

	Another averaging concerns random events of photon scattering by atmospheric inhomogeneities and statistical  averaging over different runs. When propagated over long distances, light acquires the properties of Gaussian statistics, which leads to  splitting of the fourth-order correlation function into two products of the second-order correlation functions:
	\begin{widetext}
			\begin{equation}\label{tryd5}
			\langle\overline{b_ {{\bf q}{+}\frac{\bf k}{2}}^\dag b_ {{\bf q}{-}\frac{\bf 	k}{2}}b_ {{\bf q}_1{+}\frac{{\bf k}_1}{2}}^\dag b_{{\bf q}_1{-}\frac{{\bf k}_1}{2}}}\rangle_t \approx\langle\overline{ b_ {{\bf q}{+}\frac{\bf k}{2}}^\dag b_ {{\bf q}{-}\frac{\bf k}{2}}}\rangle_t \langle\overline{ b_ {{\bf q}_1{+}\frac{{\bf k}_1}{2}}^\dag b_ {{\bf q}_1{-}\frac{{\bf k}_1}{2}}}\rangle_t+
			\langle \overline{b_ {{\bf q}+\frac{\bf k}{2}}^\dag  b_ {{\bf q}_1{-}\frac{{\bf 	k}_1}{2}}}\rangle_t \langle \overline{b_ {{\bf q}{-}\frac{\bf k}{2}} b_ {{\bf q}_1{+}\frac{{\bf k}_1}{2}}^\dag}\rangle_t.
		\end{equation}
	\end{widetext}
	The first term on the right-hand side after substitution in Eq.~\eqref{tryd2} is canceled together with the term $-\langle\overline{I({\bf r}_A,t)} \rangle\langle\overline{ I({\bf r}_B,t)}\rangle$. Then using Eqs. \eqref{tryd2} and \eqref{tryd5} we obtain  
	\begin{gather}
		\langle\overline{\delta I({\bf r}_A,t) \delta I({\bf r}_B,t)}\rangle=\frac 	{1}{S^2}\sum_{{{\bf q},{\bf k},{\bf q}_1,{\bf k}_1}} e^{-i{({\bf k\cdot r}_A+{{\bf k}_1\cdot \bf{r}}_B)}} \nonumber \\ \times\langle \overline{b_ {{\bf q}+\frac{\bf k}{2}}^\dag  b_ {{\bf q}_1{-}\frac{{\bf k}_1}{2}}}\rangle_t \langle \overline{b_ {{\bf q}{-}\frac{\bf k}{2}} b_ {{\bf q}_1{+}\frac{{\bf k}_1}{2}}^\dag}\rangle_t.
		\label{tryd6}
	\end{gather}
	It is seen that the last factor is not yet \textquotedblleft decoupled\textquotedblright,\, which is due to the influence of the phase diffuser. We use alternative variables to represent this term as the product of two components. New variables are defined by 
	\begin{equation}
		\begin{split}
	     & {\bf Q} = \frac 12 \left({{\bf q}+{{\bf q}_1}+\frac {{\bf k}-{\bf k}_1}{2}}\right), \\
		 & {\bf Q}_1 = \frac 12 \left({{\bf q}+{{\bf q}_1}-\frac {{\bf k}-{\bf k}_1}{2}}\right),
		\end{split}
	 	\label{tryd7}
	\end{equation}
	and 
	\begin{equation}
		\begin{split}
		 & {\bf K} ={{\bf q}-{{\bf q}_1}+\frac {{\bf k}+{\bf k}_1}{2}},\\
		 & {\bf K}_1 = {{\bf q}_1-{\bf q}+\frac {{\bf k}+{\bf k}_1}{2}}.
		\end{split}
		\label{tryd8} 
	\end{equation} 
	Then we have for the last multiplier in \eqref{tryd6}
	\begin{widetext}
		\begin{equation}
			\begin{split}
			 \langle\overline{ b_ {{\bf q}+\frac{\bf k}{2}}^\dag  b_ {{\bf 		q}_1{-}\frac{{\bf k}_1}{2}}}\rangle_t\langle \overline{b_ {{\bf q}{-}\frac{\bf k}{2}} b_ {{\bf q}_1{+}\frac{{\bf k}_1}{2}}^\dag} \rangle_t \equiv & \, \langle\overline{b_ {{\bf Q}+\frac{\bf K}{2}}^\dag  b_ {{\bf Q}{-}\frac{\bf K}{2}}}\rangle_t\langle\overline{ b_ {{\bf Q}_1{-}\frac{{\bf K}_1}{2}} b_ {{\bf Q}_1{+}\frac{{\bf K}_1}{2}}^\dag} \rangle_t \\
			 = & \, \exp\bigg [\frac{r_1^2-r_0^2}4\bigg(({\bf Q}-{\bf Q}_1)^2-\frac{({\bf K}-{\bf K}_1)^2}{4}\bigg)\bigg] \langle\overline{ b_ {{\bf Q}+\frac{\bf K}{2}}^\dag  b_ {{\bf Q}{-}\frac{\bf K}{2}}}\rangle_t\langle\overline{ b_ {{\bf Q}_1{-}\frac{{\bf K}_1}{2}} b_ {{\bf Q}_1{+}\frac{{\bf K}_1}{2}}^\dag} \rangle_t.
			 \label{tryd9}
			\end{split} 
	\end{equation}
\end{widetext}
	Here the equality of the first and third expressions is ensured by an adequate choice of the exponential multiplier. This  multiplier is obtained from the initial conditions for the product of four light amplitudes in \eqref{tryd9}:
	\begin{widetext}
		\begin{gather}  
			\exp\bigg[\frac{r_1^2-r_0^2}4\bigg(({\bf Q}-{\bf Q}_1)^2-\frac{({\bf K}-{\bf 	K}_1)^2}{4}\bigg)\bigg]=\exp\bigg[-(q^2+q^2_1)\frac{r_1^2}2-(k^2+k^2_1)\frac{r_0^2}8 \bigg] \nonumber \\ \times \exp\bigg[(Q^2+Q^2_1)\frac{r_1^2}2+(K^2+K^2_1)\frac{r_0^2}8 \bigg].
			\label{sor} 
		\end{gather} 
	\end{widetext} 
	
	It follows from Eq.~\eqref{tryd9} that two pairs of photon amplitudes are represented by two pairs of independent variables. This circumstance simplifies further consideration. Further simplification concerns equation \eqref{tryd9} in which the term $({\bf K}-{\bf K}_1)^2/4$ can be neglected in the case of long distance propagation.
	Also, it is useful to represent the latter factor as
	\begin{equation}\label{sor1}    
		\langle\overline{ b_ {{\bf Q}_1{-}\frac{{\bf K}_1}{2}} b_ {{\bf Q}_1{+}\frac{{\bf K}_1}{2}}^\dag} \rangle_t=\delta_{{\bf K}_1=0}+\langle\overline{ b_ {{\bf Q}_1{+}\frac{{\bf K}_1}{2}}^\dag b_ {{\bf Q}_1{-}\frac{{\bf K}_1}{2}}}\rangle_t.
	\end{equation}
	To complete this part of the analysis, we will rewrite the factor that describes the spatial distribution in the new variables: 
	\begin{gather}
		e^{-i{({\bf k\cdot r}_A+{{\bf k}_1\cdot {\bf r}}_B)}} =\exp{\{-i[({\bf K}+{\bf K}_1)\cdot \frac{{\bf r}_A+{\bf r}_B}2]\}} \nonumber \\ \times\exp{\{-i[({\bf Q}-{\bf Q}_1)\cdot({\bf r}_A-{\bf r}_B)]\}}.
			\label{sor2}
	\end{gather}
Finally, using the definition of the PDF and equations \eqref{tryd6}, \eqref{tryd9}, \eqref{sor1}, the two-dimentional correlation function  of photon density is presented by
	\begin{widetext}
		\begin{gather}
			\langle\delta I({\bf r}_A) \delta I({\bf 	r}_B)\rangle=\frac{1}{\pi(r_0^2-r_1^2)}\exp{\bigg[-\frac{({\bf r}_A-{\bf r}_B)^2}{r_0^2-r_1^2}\bigg]}\langle I\big(\frac{{\bf r}_A+{\bf r}_B}{2}\big)\rangle+\sum_{{\bf q},{\bf q}_1}\langle f\big(\frac{{\bf r}_A+{\bf r}_B}{2},{\bf q}\big)\rangle\langle f\big(\frac{{\bf r}_A+{\bf r}_B}{2},{\bf q}_1\big )\rangle \nonumber \\ \times \exp\left[i({\bf q}_1-{\bf q})\cdot({\bf r}_A-{\bf r}_B)-\frac{1}{4}({\bf q}_1-{\bf q})^2(r_0^2-r_1^2)\right].
			\label{sor3}
		\end{gather} 
	\end{widetext}
	
	Equation \eqref{sor3} is similar to equation \eqref{trynad} and considers the influence of the phase diffuser. The diffuser can dramatically change the characteristics of the beam. First of all, the shot-noise term is no longer delta-correlated but instead is described by the finite-width function of $({\bf r}_A-{\bf r}_B)$. This function is equal to 
	\begin{equation}\label{sor4}
		\frac{1}{\pi(r_0^2-r_1^2)}\exp\bigg[-\frac{({\bf r}_A-{\bf r}_B)^2}{r_0^2-r_1^2}\bigg]
	\end{equation}   
	and tends towards $\delta\big({\bf r}_A-{\bf r}_B\big)$, when  $r_1\rightarrow r_0$ - as in Eq.~\eqref{trynad}.

	Also, the  factor  $e^{-({\bf q}-{\bf q}_1)^2(r_0^2-r_1^2)/4}$, describing the reduction of quasiclassical fluctuations, appears in the last term of \eqref{sor3}. This factor is negligibly small when $r_1 \ll r_0$, so fluctuations are suppressed. [For this, it is sufficient the inequality $\langle q^2\rangle r^2_0\gg1$ to be satisfied]. In contrast, this factor tends to unity when $r_1\rightarrow r_0$ as it should be without a phase diffuser. In the moderate case, this factor is estimated as $\exp\big[-\langle q^2\rangle(r_0^2-r_1^2)/4\big].$  
	This value can be used to estimate the scintillation index $\sigma ^ 2 $ in the center of the beam, where $ r_A = r_B = 0 $. 

\section{Discussion and conclusion}
	\label{sec:Dis}
 
 	The use of the Gaussian statistics reduces the theoretical studies of beam wandering and the effect of the phase diffuser to the analytical integration of average PDF values. The simplicity of these distributions, obtained in \cite{bb}, facilitates the analytical study of the physical properties of beams. 

	A significant difference in the time dependences of beam wandering in quantum and classical pictures of propagation is shown in Sec.~\ref{sec:Wander}.
	The difference is explained by the fragmentation of the beam, up to a single-photon level, caused by multiple collisions with atmospheric turbulent vortices.  Mathematically, this means the loss of correlation between different parts of the beam in the case of long propagation distances. The contribution of the classical part in wandering decreases. The opposite tendency is observed for the quantum part.  

	The mechanism of formation of partial coherence using a phase diffuser is explained in section \ref{sec:Dif}. To represent the correlation function of fluctuations in terms of the average PDF, the special \textquotedblleft splitting\textquotedblright procedure is developed. As a result, the general expression \eqref{sor3} describing the influence of the phase diffuser is obtained.  Without the phase diffuser, this expression reduces to that, given by equation \eqref{trynad}. The modified version allows obtaining the FM using the average value of the PDF. This expression is used in Sec.~\ref{sec:Dif} to estimate the effect of the phase diffuser. 

	The phase diffuser modifies the shot noise by transforming the spatial delta-function correlated factor into a smooth function. The maximum value and width of this function depend on the diffuser parameter $\langle g^2\rangle$. 

	It should be noted that our results are obtained within the paraxial approximation, which may break down at very long distances.  

	The motivation of the authors is to study the fluctuation properties of beams due to the high urgency of the problem.  We fully support the opinion of the authors of \cite{and2}, who stated that \textquotedblleft optical scintillation is considered one of the \textbf{most important atmospheric effects that must be fully understood, as they ultimately determine the limitations of optical system performance\textquotedblright\,.}

\section*{Acknowledgments} 

	This work was carried out under the auspices of the Institute of Physics of the National Academy of Sciences of Ukraine, Project IF-2021-1 part 1. The authors thank E.~Stolyarov and A.~Negriyko for useful discussions and comments.

\def\bibsection{\section*{\refname}} 
\bibliographystyle{iopart-num}	
\bibliography{bibliography}

\end{document}